\documentclass[useAMS,usenatbib]{mn2e}

\title[Uncertainties on atomic data. A case study: \ion{N}{iv}]
{Uncertainties on atomic data. A case study: \ion{N}{iv}}
\author[Del Zanna et al.]{G. Del Zanna$^{1}$\thanks{E-mail: gd232@cam.ac.uk},
L. Fern\'{a}ndez-Menchero$^{2}$,  N.~R. Badnell$^{3}$\\
$^{1}$ DAMTP, Centre for Mathematical Sciences, University of Cambridge, Wilberforce Road, Cambridge CB3 0WA, UK \\
$^{2}$ CTAMOP, Queen's University of Belfast, Belfast, Northern Ireland, UK \\
$^{3}$ Department of Physics, University of Strathclyde, Glasgow G4 0NG, UK \\
}
\date{Submitted to MNRAS  }

\pagerange{\pageref{firstpage}--\pageref{lastpage}} \pubyear{2016}

\DeclareMathAlphabet{\mathsc}{OT1}{cmr}{m}{sc}
\def\testbx{bx}%
\DeclareRobustCommand{\ion}[2]{%
\relax\ifmmode
\ifx\testbx\f@series
{\mathbf{#1\,\mathsc{#2}}}\else
{\mathrm{#1\,\mathsc{#2}}}\fi
\else\textup{#1\,{\mdseries\textsc{#2}}}%
\fi}

\usepackage{graphicx}
\usepackage{txfonts}

\usepackage{epsfig}
\usepackage{natbib}
\usepackage{color}

\begin{document}

\label{firstpage}
\maketitle

\begin{abstract}
We consider three recent large-scale 
calculations for the radiative and electron-impact excitation data of  \ion{N}{iv},
carried out with different methods and codes. The scattering calculations
 employed the relativistic  Dirac $R$-matrix (DARC) method,
 the  intermediate coupling frame transformation (ICFT) $R$-matrix method,
and the  B-spline $R$-matrix (BSR) method.
These are all large-scale scattering calculations with well-tested
and sophisticated codes, which use the same set of target states.
One concern raised in previous literature is related to the increasingly large
discrepancies in the effective collision strengths 
between the three sets of calculations for increasingly weak and/or high-lying transitions. 
We have built three model ions and calculated the intensities of all the 
main spectral lines in this ion. We have found that, despite such large differences,
 excellent agreement (to within $\pm$~20\%)
exists between all the spectroscopically-relevant line intensities.
This provides confidence in  the reliability of the calculations 
for plasma diagnostics. 
We have used the differences in the radiative and excitation rates
amongst the three sets of calculations to obtain a measure of the uncertainty in
each rate. Using a Monte Carlo approach, we have shown how these 
uncertainties affect the main theoretical ratios which are used to measure 
 electron densities and temperatures. 
\end{abstract}


\section{Introduction }

With the advances in computational power, an increasing number of 
large-scale atomic calculations for ions of astrophysical 
importance have become available in the past few years. 
Within several communities, there is now awareness of the 
importance of the accuracy of atomic calculations for astrophysical applications. 
The IAEA has organised a workshop on uncertainties 
of atomic data, and some guidance is provided in e.g. \cite{chung_etal:2016}. 
Within the solar community, 
\cite{guennou_etal:2013} used some general estimates of available 
rates, to assess uncertainties on the  temperature distribution
of the solar  plasma, obtained by combining spectroscopic 
observations with  atomic data. The resulting uncertainties were large.
A much better approach is to assess in some way the uncertainty in each
single rate, by e.g. comparing theory with experiment or 
results of different calculations. One of us (GDZ) developed such an approach,
and used comparisons of two calculations to assess  uncertainties associated 
with  line ratios used to measure electron densities from 
\ion{Fe}{xiii} \citep{yu_etal:2018}.
A general overview of some of the uncertainties in  atomic 
data is presented in \cite{delzanna_mason:2018}.

For most astrophysical plasmas that are collisional (i.e. not photoionised),
the main rates affecting the spectral lines of
an ion  are those of spontaneous decay (A-values), and those
of collisional excitation by electron impact. 
In this paper, we focus on the latter, considering that 
they are normally the most complex  ones to calculate accurately. 
The main quantity is  the effective collision strength $\Upsilon$
of a transition, which is the rate obtained from the adimensional 
cross-section (the collision strength) assuming a Maxwellian 
distribution of the electrons.

Historically, large discrepancies in the effective collision strengths calculated
 with different approximations and codes were present. 
However, with the advances in computational power,
results have generally converged for low-lying transitions. 
However, in the recent literature, several cases have now appeared 
where differences of up to  one or two orders of magnitude for weak and/or high-lying transitions
have been found. This  clearly raises concerns on the reliability of any of such 
calculations, and their effects on diagnostic applications.

Generally, the  effective collision strengths  to the lower levels of an ion 
agree to within  $\pm$20\%, although in a few cases they can differ significantly.
One such example are the calculations for the coronal 
\ion{Fe}{xi} ion, where the values calculated by \cite{delzanna_etal:10_fe_11}
were  generally in good agreement (within $\pm$20\%) with those  calculated 
by \cite{aggarwal_keenan:2003b},
with the exception of a few amongst the strongest ones, 
where large differences were present. They occurred for 
levels which have a strong spin-orbit interaction, so the calculations are
very sensitive to the atomic structure.

We are concerned here with those cases where the overall 
results are significantly different. There are several examples in the
 literature, some of which have been recently reviewed by 
\cite{aggarwal:2017Atoms}. For example, 
\cite{aggarwal_keenan:2014} used the Dirac Atomic $R$-matrix
Code (DARC) of P. H. Norrington and I. P. Grant to 
calculate  effective collision strengths for the important coronal 
\ion{Fe}{xiv}. They found large discrepancies with the 
results  obtained by 
\cite{liang_etal:10_fe_14} with the Intermediate Coupling Frame Transformation (ICFT) 
$R$-matrix method \citep{griffin_etal:98}.
The \cite{aggarwal_keenan:2014}  calculations adopted 
much smaller configuration-interaction (CI) and 
close-coupling (CC) expansions than the previous study, so significant
differences are to be expected. 
Indeed,  \cite{delzanna_etal:2015_fe_14} carried out a new ICFT calculation
with the same CC/CI expansions as that one 
adopted by  \cite{aggarwal_keenan:2014} and found excellent agreement
between the DARC and ICFT results. The main differences between the smaller DARC 
and the larger ICFT  calculation was the CC expansion used.

A similar case concerns \ion{Al}{x}.  \cite{aggarwal_keenan:2015}
carried out a  DARC 98-levels calculation on this ion, showing 
significant differences with the results obtained by \cite{be-like} 
with a much larger (238-level) ICFT calculation. 
As in the  \ion{Fe}{xiv} case, \cite{be-like_rebuttal} carried out 
ICFT and Breit--Pauli $R$-matrix calculations with the same  target (98-levels) 
adopted by  \cite{aggarwal_keenan:2015}, and the results  
 were shown to agree  closely. 
Various comparisons were also provided, showing how both the choices of 
CC and  CI expansions can significantly affect the collision strengths. 

On a side note, we stress that  it has been  
shown in the literature that when the atomic structure of an ion is similar,
and the same CI/CC expansions are adopted, results of the 
DARC and ICFT are very similar 
(see, e.g. \citealt{liang_badnell:2010_ne-like,liang_etal:09_na-like,badnell_ballance:2014}).
Clearly, many other issues affect the final results, such as 
the  energy resolution and the threshold positions, but are often of 
less importance, as we have discussed e.g. in the  \cite{badnell_etal:2016} review. 

There are also reported differences in collision strengths obtained with other codes. 
For example, \cite{aggarwal_keenan:2017_mg_5} carried out a
DARC $R$-matrix calculation on \ion{Mg}{v} on 86 target states and found large, 
order of magnitude  differences for weak and/or high-lying transitions
with two  previous results. One was an earlier (and smaller)
ICFT $R$-matrix calculation carried out by 
\cite{hudson_etal:2009_mg_5}. The other one, by 
\cite{tayal_sossah:2015_mg_5}, was on the same 86 target states
but  used a completely different approach, the B-spline $R$-matrix (BSR) method
\citep[see, e.g. ][ for details]{zatsarinny:2006,zatsarinny_bartschat:2013}.

In a recent study, \cite{wang_etal:2017_mg_5} carried out two sets of 
BSR calculations, one  with the same 86 target states as
in \cite{tayal_sossah:2015_mg_5}, but with a 
 more accurate representation of the target structure; 
the other one was a much larger calculation with 316 states. Close agreement 
with the previous BSR calculations was found when the smaller 
calculation was considered. Significant increases in the 
 collision strengths  of  the weaker transitions was however  found when the 
much larger calculation was considered. 
We note that such results are common and are mostly due to extra resonances, 
and coupling in general, which can  increase the cross-sections for weak transitions.  
Differences with the DARC results were found.

For the present study we have chosen to consider the  Be-like  \ion{N}{iv},
as  effective collision strengths  obtained with the 
 DARC, ICFT, and BSR $R$-matrix codes are now available. They were obtained 
 with the same set of target states, hence are directly comparable.
\cite{aggarwal_keenan:2016}  
  carried out a  DARC scattering calculation for \ion{N}{iv},
showing order of magnitude discrepancies for many weak and/or high-lying transitions with the 
values calculated previously by \cite{be-like} with the ICFT $R$-matrix codes
[hereafter ICFT results]. 
The DARC calculations adopted  the same set of configurations for the CI/CC expansions as 
those used by \cite{be-like}. 
Therefore, in principle one would expect good agreement,
at least for those transitions where slight differences in the 
atomic structure (which are always present) are not significant.

To  shed light into this issue, 
\cite{fernandez-menchero_etal:2017}   recently carried out a 
large-scale calculation with the  BSR codes. 
Good agreement between all calculations 
was found  for the strong transitions within the low-lying states.
Significant increasing differences with both 
the ICFT and DARC results were found for the increasingly weaker and/or
higher transitions. The differences are attributable to the inherent lack of
convergence in the target configuration interaction expansion and/or
the collisional close-coupling expansion in all three calculations,
which increasingly affects the weaker and/or higher-lying transitions.
The convergence study by \cite{be-like_rebuttal} illustrates this point.

In this paper, we focus on two important aspects: 
1) we show that the large differences have negligible effects for 
astrophysical modelling; 
2) we  use  the differences as a measure of 
the uncertainty in the rates, and provide a measure of the 
uncertainty in derived quantities such as electron densities.

It is well-known that  a few of the strongest  \ion{N}{iv} lines 
are useful diagnostics for astrophysical plasma
(e.g. nebulae and the solar corona), see e.g. \cite{dufton_etal:1979} 
 Some have also been used in laboratory plasma (tokamaks). 

\begin{figure}
\centerline{\epsfig{file=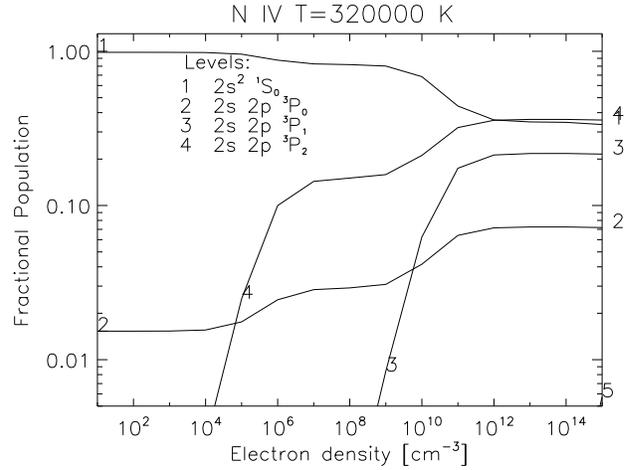, width=6.5cm,angle=90 }}
\caption{Relative populations of the levels of \ion{N}{iv}.}
\label{fig:pop}
\end{figure}

\section{Modelling spectral line intensities}

\begin{figure}
\centerline{\epsfig{file=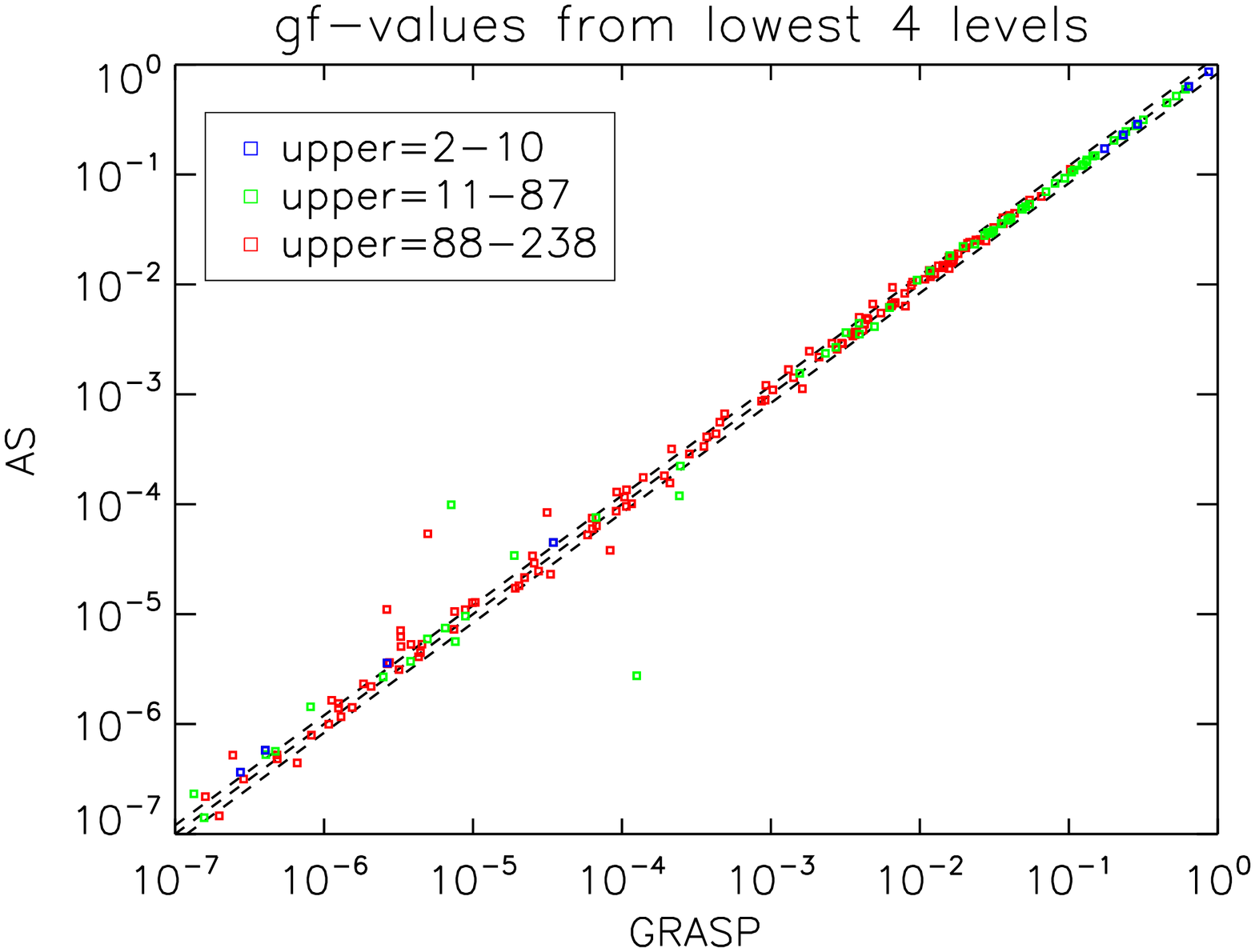, width=6.0cm,angle=90 }}
\centerline{\epsfig{file=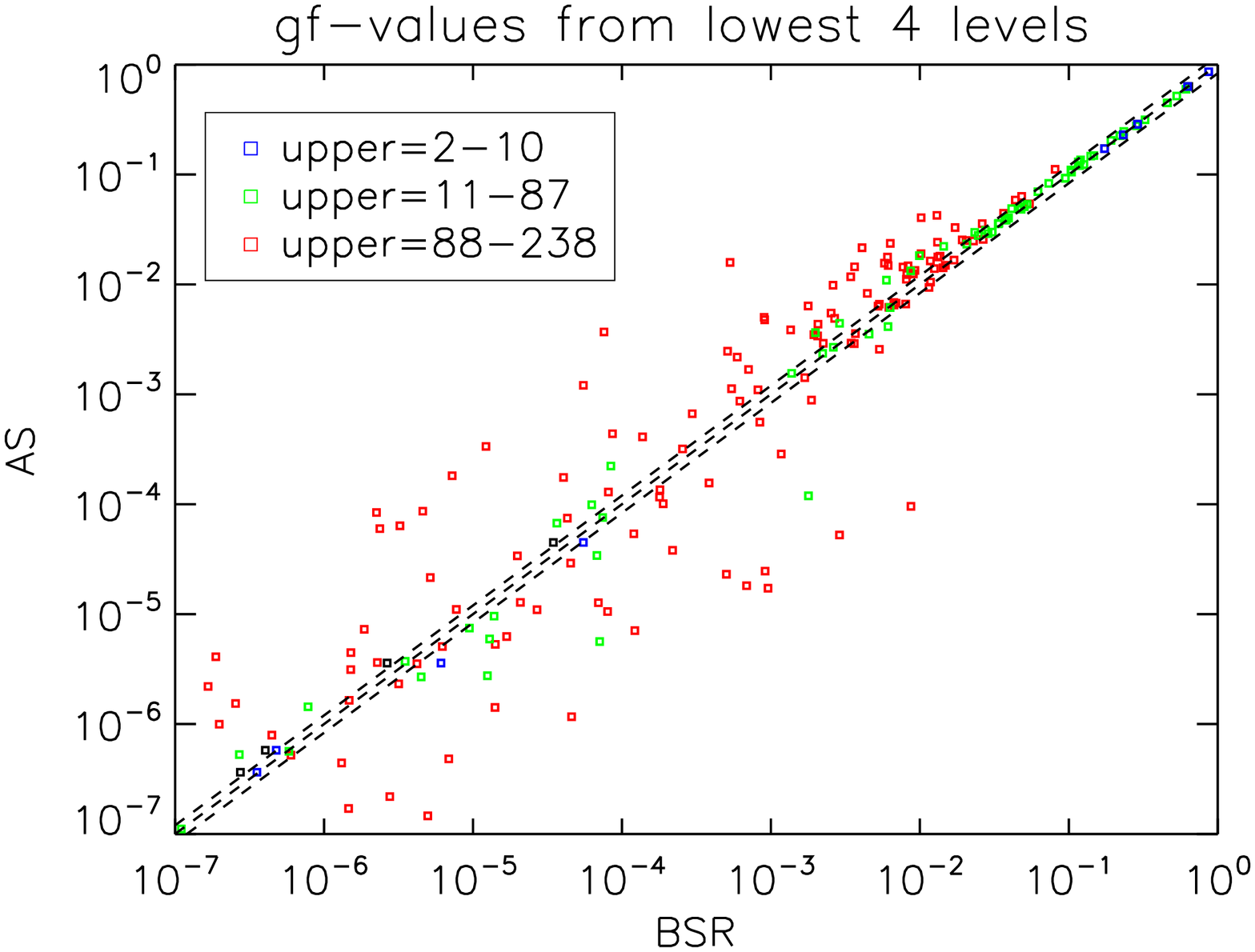, width=6.0cm,angle=90 }}
\centerline{\epsfig{file=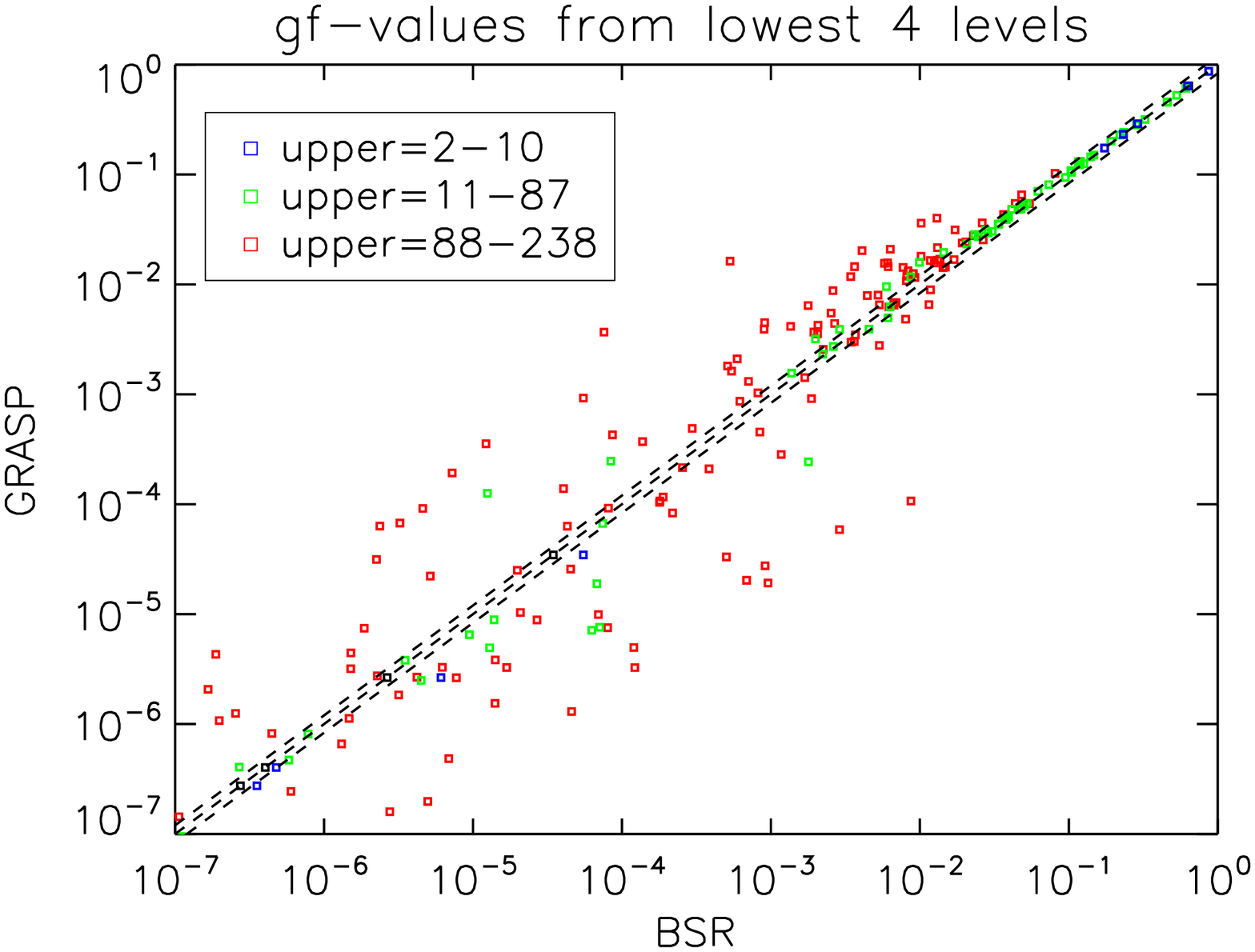, width=6.0cm,angle=90 }}
\caption{Comparisons of gf-values as calculated with the GRASP, AS, and BSR
codes for all transitions from the metastable levels. 
Dashed lines show $\pm$ 20\%. }
\label{fig:gf}
\end{figure}

\begin{figure}
\centerline{\epsfig{file=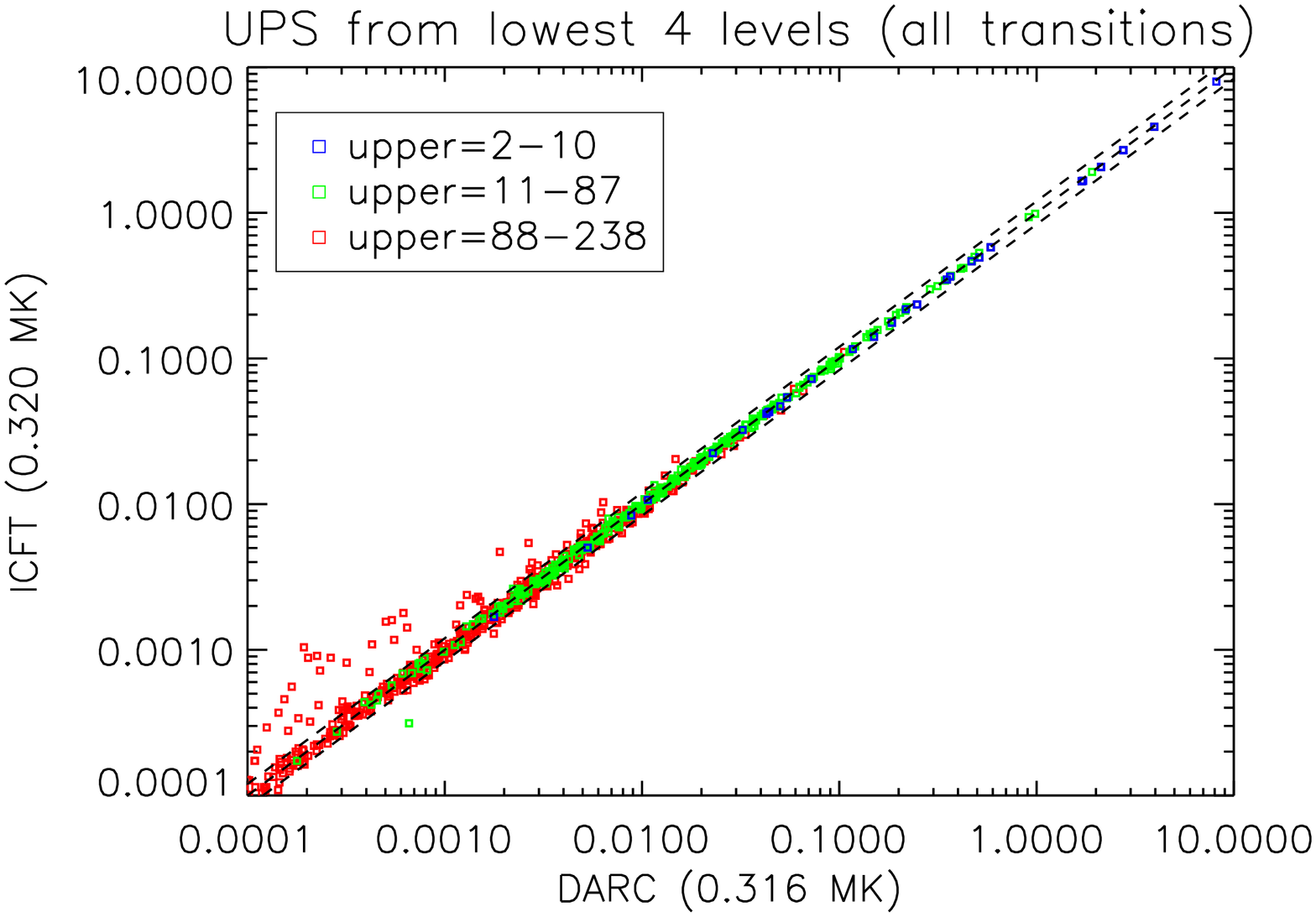, width=6.0cm,angle=90 }}
\centerline{\epsfig{file=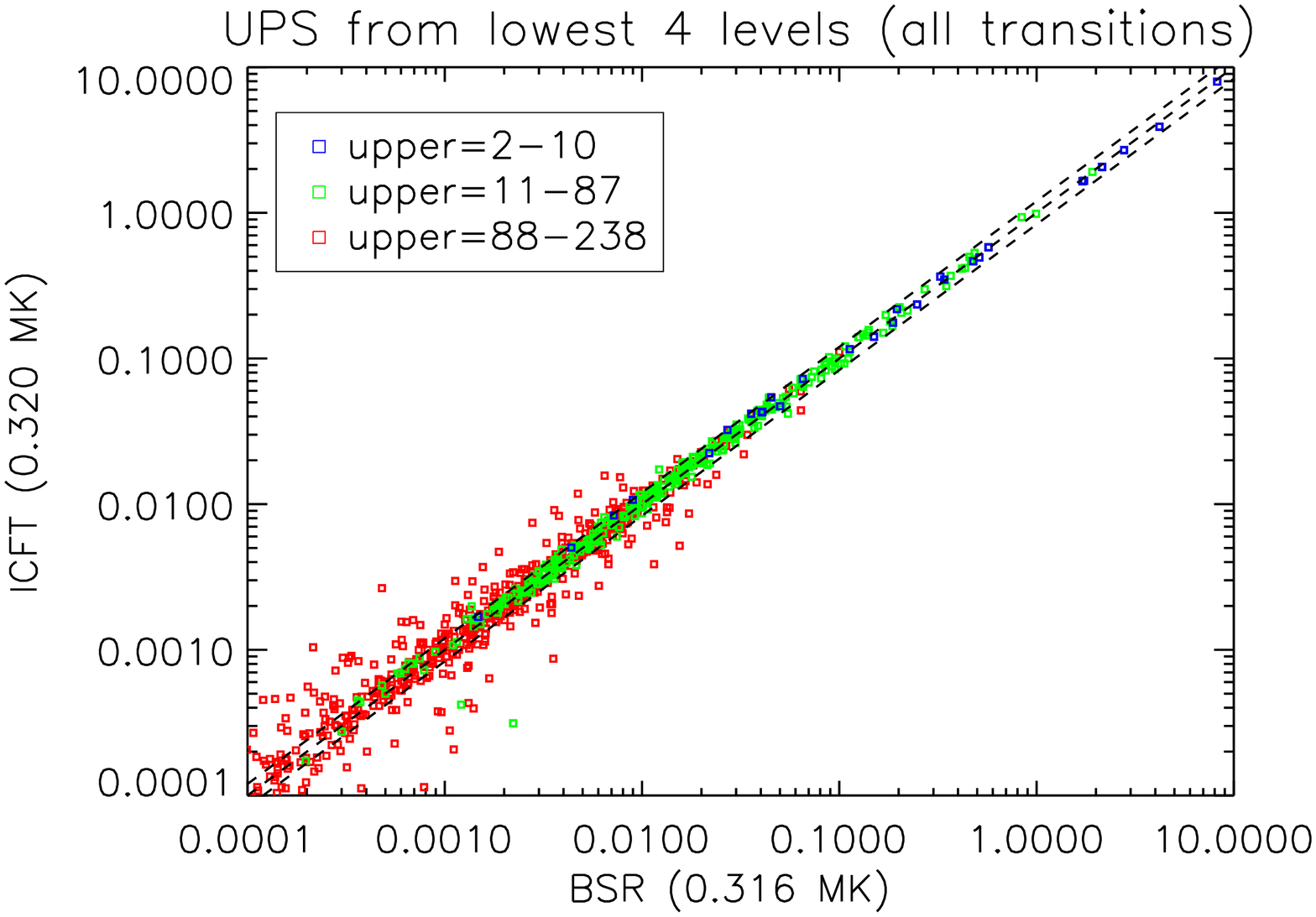, width=6.0cm,angle=90 }}
\centerline{\epsfig{file=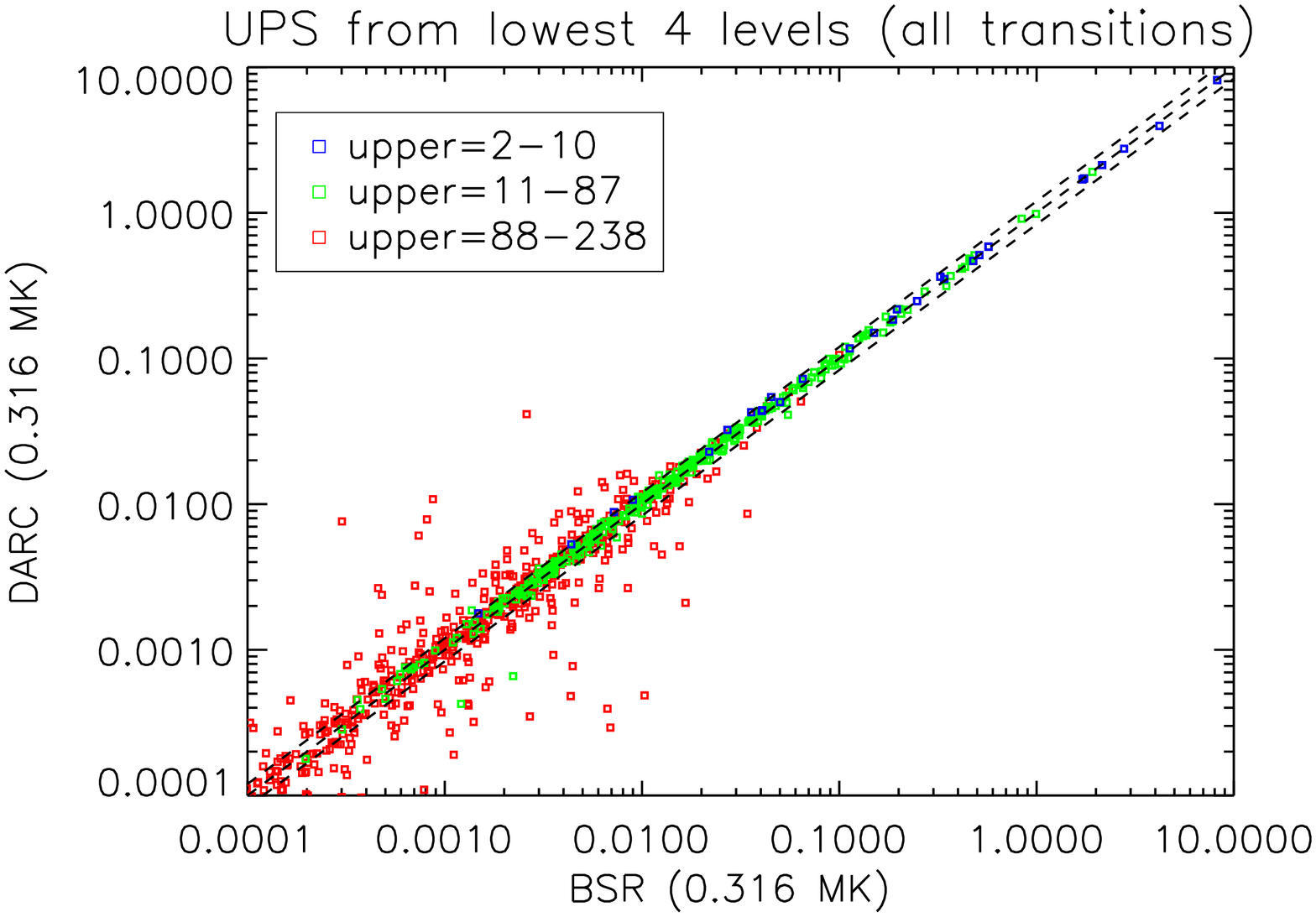, width=6.0cm,angle=90 }}
\caption{Comparisons of effective collision strengths $\Upsilon$ (UPS)
near ion peak abundance as calculated with the DARC, ICFT, and  BSR codes,
for all  transitions from the metastable levels.}
\label{fig:ups}
\end{figure}

\subsection{Atomic data}

\cite{aggarwal_keenan:2016} carried out two sets of atomic structure calculations using 
the GRASP  (General-purpose Relativistic Atomic Structure Package) code, 
originally developed by \cite{grant_etal:1980} and then revised by P. H.
Norrington. We consider here only the larger one, which the authors
labeled as GRASP2. This considered configuration interaction (CI) 
producing the same set of 238 fine-structure levels adopted by \cite{be-like}.
They included  a relatively complete set of configurations  up to 
principal quantum number $n=5$, plus 72 levels arising from 
$n=6,7$ configurations, which was added  by \cite{be-like} 
to improve the structure for the lower levels.

\cite{be-like}, on the other hand,  used 
the {\sc autostructure} (AS) program \citep{badnell:11} and 
 radial wavefunctions calculated 
in a scaled Thomas-Fermi-Dirac-Amaldi statistical model potential.
As shown by \cite{aggarwal_keenan:2016}, some differences in the energies of the two 
calculations are present, although not apparently large enough 
to expect large differences in the results of the scattering 
calculations. 

For the scattering calculations, \cite{aggarwal_keenan:2016}  used the 
relativistic DARC program. For the comparisons shown here we only consider 
the results of their `DARC2' calculation, which included in the CC
expansion the same set of 238 levels of the CI. 

\cite{be-like} instead used a set of codes and methods some of which 
originated from the Iron Project, and are described 
in e.g. \cite{hummer_etal:93} and \cite{berrington_etal:95}.
The $R$-matrix inner region calculation was 
in $LS$-coupling and included  both mass-velocity and Darwin relativistic energy
corrections. For the outer region, the ICFT method 
was applied to the  $LS$-coupled $K$-matrices calculated with the STGF code
(Badnell and Seaton, unpublished).
Collision strengths were `topped-up' to infinite partial waves
following \cite{burgess:74,badnell_griffin:01}.
Finally, the collision strengths were extended to high
 energies by interpolation using  the 
appropriate high-energy limits in the 
 \cite{burgess_tully:92} scaled domain.
The high-energy limits were calculated with {\sc autostructure}
following \cite{burgess_etal:97} and \cite{chidichimo_etal:03}.

As we mentioned, \cite{fernandez-menchero_etal:2017}
 carried out a  large-scale calculation with the B-spline $R$-matrix codes.
They included all the valence configurations $\{2s,2p,3s,3p,3d\}\,nl$ of N$^{3+}$.
The outer valence-electron $nl$ wave function was expanded in a basis set of 134 B-splines of order 8.
The BSR calculation was limited to the total angular momenta $J=0-6$, obtaining a total of 1400  levels,
including bound and continuum. Of these 1400 levels calculated, 238 were included in the later CC expansion.
They used the same outer region STGF code, but now in $jK$-coupling.

We used the collision strengths and A-values
published by  \cite{fernandez-menchero_etal:2017} to calculate the level population
for this ion at the temperature of maximum abundance in 
ionization equilibrium. We used the codes available within 
the CHIANTI package \citep{dere_etal:97,delzanna_etal:2015_chianti_v8}.
Figure~\ref{fig:pop} shows that for any astrophysical density,
and for any plasma laboratory density below 10$^{15}$ cm$^{-3}$,
the levels that drive the population of all the levels in the ion
are the ground state (2s$^2$ $^1$S$_{0}$) and the  three metastable levels, 
from the 2s 2p $^3$P. 
This is an important issue: the intensities of the spectral lines
are directly proportional to the populations of the upper levels, which in 
turn are driven solely by the collision rates from these four lower levels.
All the rates from the other levels are irrelevant for the modelling.

We therefore looked at the gf values (weighted oscillator strengths)
 of all the transitions from these four levels,
as calculated by \cite{aggarwal_keenan:2016}  with GRASP and \cite{be-like} with AS.
They are shown in  Figure~\ref{fig:gf} (top).
With very few notable exceptions, there is excellent agreement,
to within $\pm$ 20\%, for all the transitions, especially the strong ones.
On the other hand, significant differences (over  $\pm$ 20\%) with the BSR calculated values
are present for the weaker transitions, as shown in the two lower plots of 
 Figure~\ref{fig:gf}.

Such differences become even more evident when 
all the transitions are considered, as shown in Fig.2
of \cite{fernandez-menchero_etal:2017}. 
These differences
result directly from the different method used in the atomic
structure calculation. The BSR calculations adopted a 
multi-configuration Hartree-Fock (MCHF) expansion which included the 
continuum in the form of pseudo-orbitals, 
and where the radial functions for the outer
valence electron were expanded in a B-spline basis.
This  generated different nonorthogonal sets of one-electron orbitals 
for each target state and the continuum, and therefore led to a 
more extended CI expansion, compared to the GRASP and AS results.
In particular, CI included configuration mixing with
the additional [3s, 3p, 3d]nl bound states, as well as the
interaction with the continuum. 
One would expect that the use of a more complete basis set in the BSR calculations
would provide a better atomic structure.  
Indeed, as shown in  \cite{fernandez-menchero_etal:2017}, the resultant 
level energies are the
ones closest to the observed values (as available in NIST or CHIANTI).

The comparison of the  $\Upsilon$ as calculated with the DARC, ICFT, and  BSR codes,
for the same set of transitions, at temperatures close
to ion peak abundance, are shown in  Figure~\ref{fig:ups}.
Excellent agreement (to within $\pm$ 20\%) between the DARC and ICFT 
is found for all transitions, with a 
few cases belonging to the higher levels 
(2p 4$l$ and most of the $n=6,7$, above level No.~87) which deviate by 
only about 30\%. 
 As in the case of the gf values, larger 
differences are found with the BSR values. 
This is expected as the high-temeprature limits of the effective collisions
strengths are directly related to the gf values.

The larger  deviations occur for the transitions to higher levels,
which are mostly  forbidden.
Such variations are  quite typical for weak forbidden lines, which are very sensitive to 
a number of issues, such as cancellation effects, the positioning
of the resonances, etc.
Much larger variations are present if one considers all transitions from all 
levels, as shown by \cite{aggarwal_keenan:2016}  and  \cite{fernandez-menchero_etal:2017}.
 However, as we pointed out, they would not
have any effect for the modelling. 

The question is whether the  30--40\% variations in the 
forbidden lines have any significant effect on the level population
for this ion. To assess this, we have build three ion models and 
solved the level population.

\subsection{Level population and line intensities}

\begin{figure}
\centerline{\epsfig{file=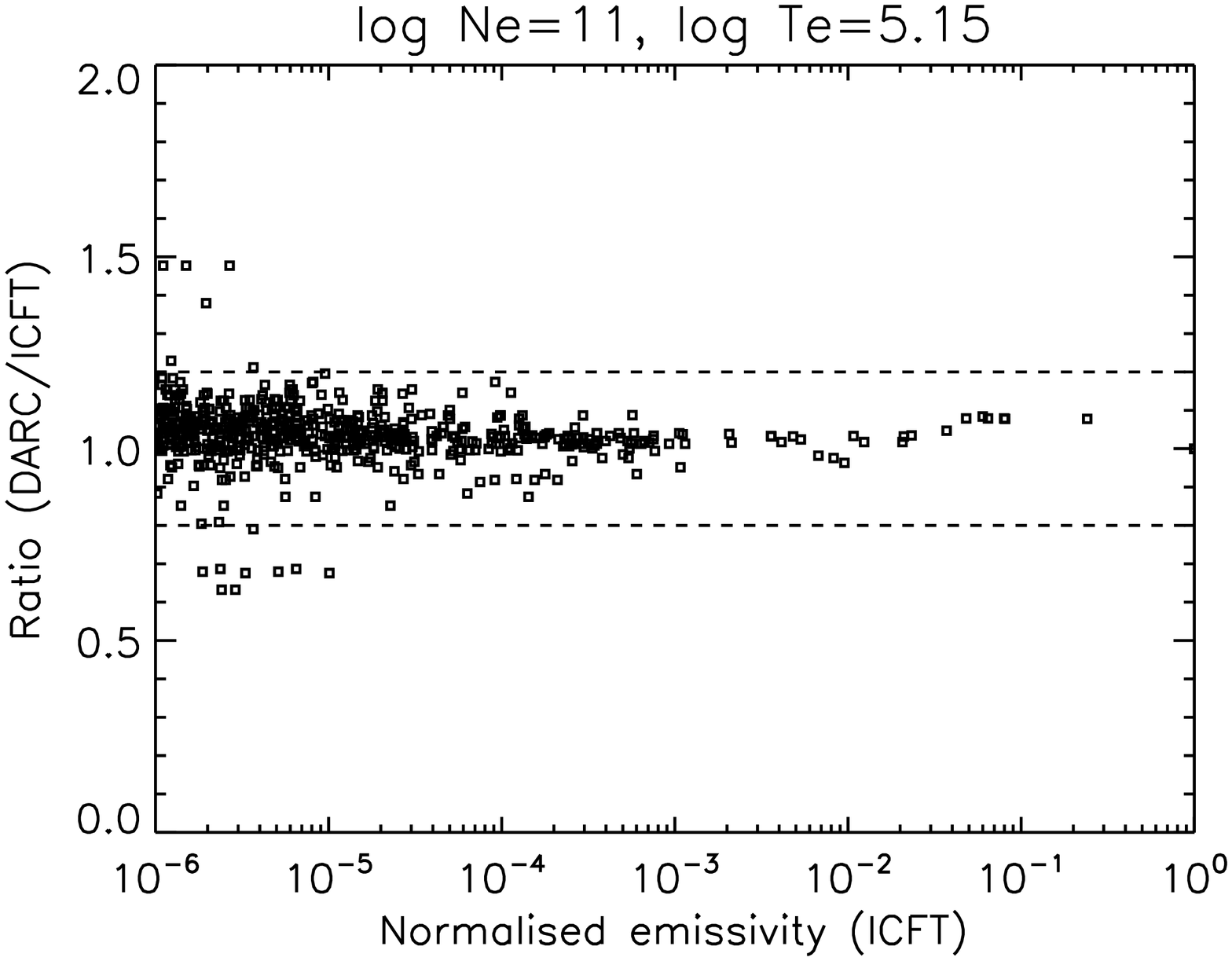, width=6.0cm,angle=90 }}
\centerline{\epsfig{file=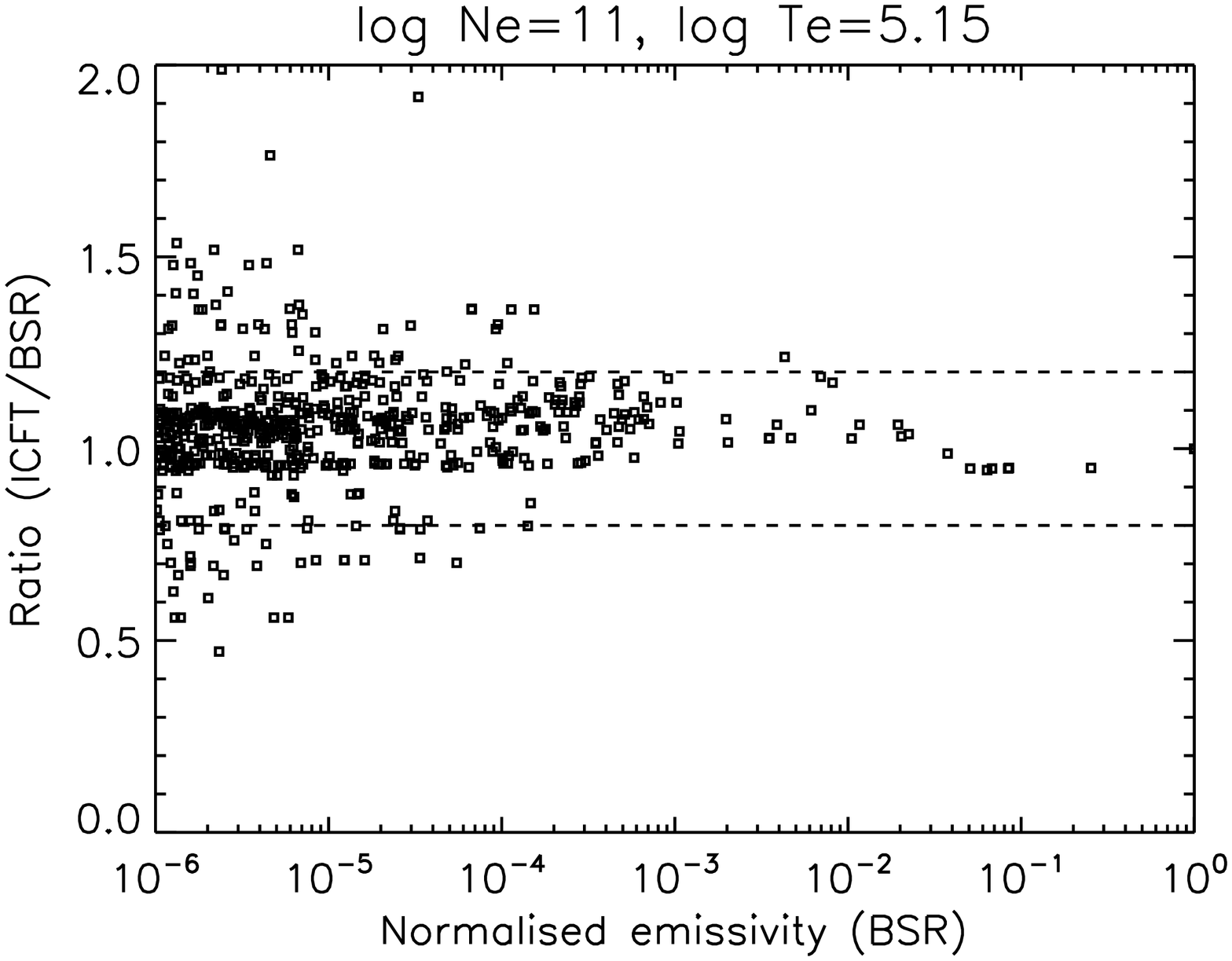, width=6.0cm,angle=90 }}
\centerline{\epsfig{file=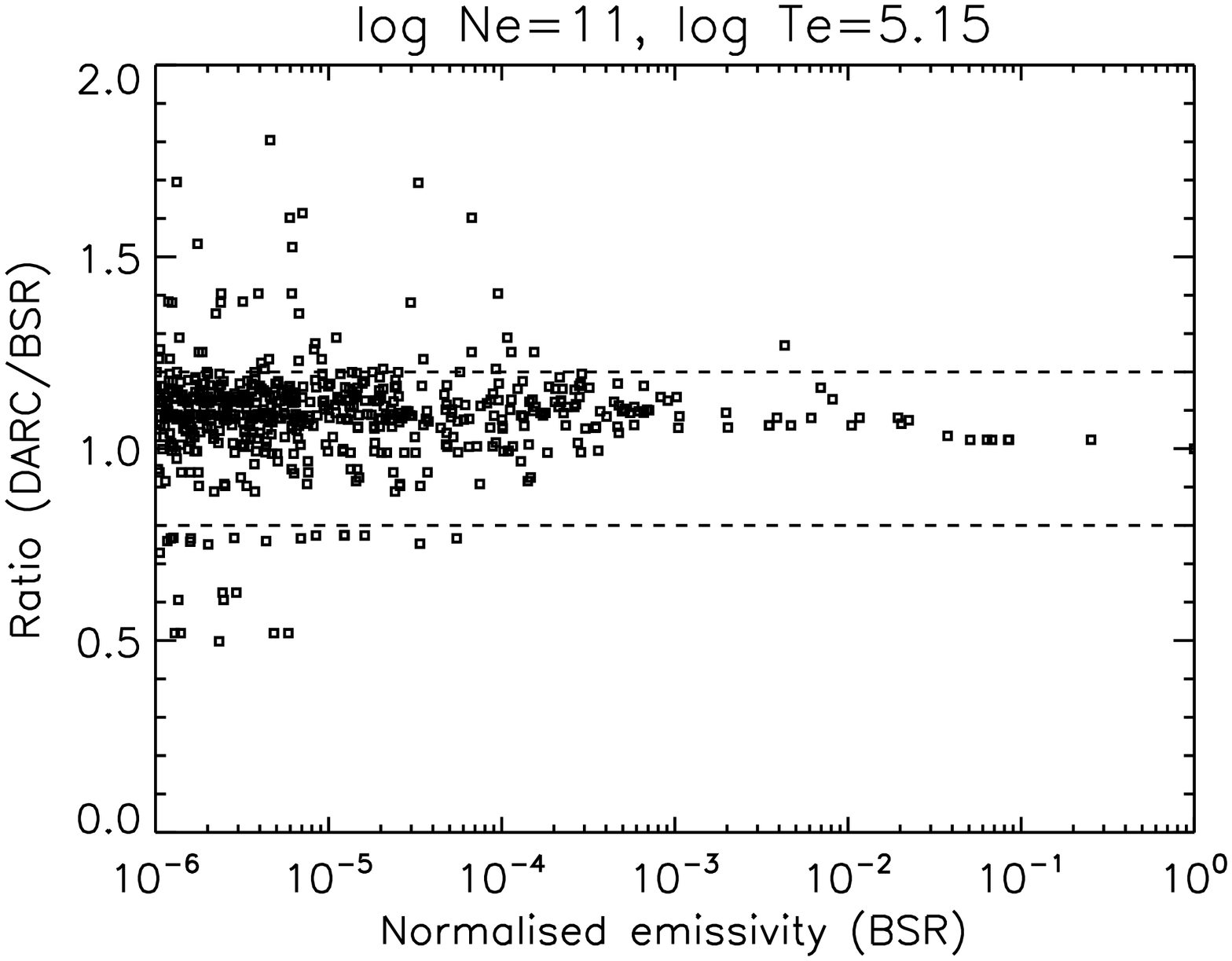, width=6.0cm,angle=90 }}
\caption{Ratios of all the  spectral line intensities, calculated with the three
model ions, based on the ICFT, DARC, and BSR $\Upsilon$ values,
as a function of their normalised intensities
(i.e. relative to the strength of the resonance line).
Dashed lines indicate $\pm$20\%.}
\label{fig:comp_emiss}
\end{figure}

Within an atomic database such as CHIANTI \citep{delzanna_etal:2015_chianti_v8}, 
one typically merges the $\Upsilon$ from a calculations with ad-hoc, normally more
accurate, radiative data (A-values) obtained by a completely 
different calculation. 
Among the three calculations we consider, as we have mentioned the
BSR one has the best atomic structure so we have adopted 
the BSR  A-values, and built three
model ions, with the BSR, ICFT and DARC effective collision strengths.
We had to switch the indexing of several levels in the 
ICFT and DARC calculations, for a meaningful comparison.

All the excitations between all 238 levels were retained, although 
as we pointed out, only those from the lowest four levels are needed
to model plasma emission below 10$^{15}$ cm$^{-3}$.
We have used the CHIANTI codes to calculate the line emissivities,
finding the level populations by including the proton rates as
available in CHIANTI v.8.

Since for diagnostic application one is interested in relative ratios,
we have considered the spectral line emissivities,  normalised
to the intensity of the strongest resonance line, the 
2s$^2$ $^1$S$_{0}$--2s 2p $^1$P$_{1}$ at 765~\AA.

Figure~\ref{fig:comp_emiss} shows the ratios of the spectral line intensities
as calculated with the ICFT,  DARC, and BSR collision strengths. 
The intensities have been calculated at ion peak abundance (log $T$[K]=5.15) and 
at an  electron density  of  10$^{11}$ cm$^{-3}$, close to the 
value expected in a solar active region.

Despite the order of magnitude differences in some of the collision strengths, 
Figure~\ref{fig:comp_emiss} clearly shows that 
there is an excellent agreement, to 
within $\pm$ 20\%, for all the spectroscopically relevant lines
which are within 4 orders of magnitude the brightest line. 
Larger differences, but still within about 50\%, are present for 
all the other
extremely weak and spectroscopically unobservable lines.
Interestingly, agreement improves when the models based on the 
 ICFT and DARC collision strengths are compared directly.
In some respects, the ICFT and DARC calculations are based on 
atomic structure calculations that are different but of similar 
accuracy.

\section{Estimating uncertainties on the main diagnostic ratios}

\begin{figure}
\centerline{\epsfig{file=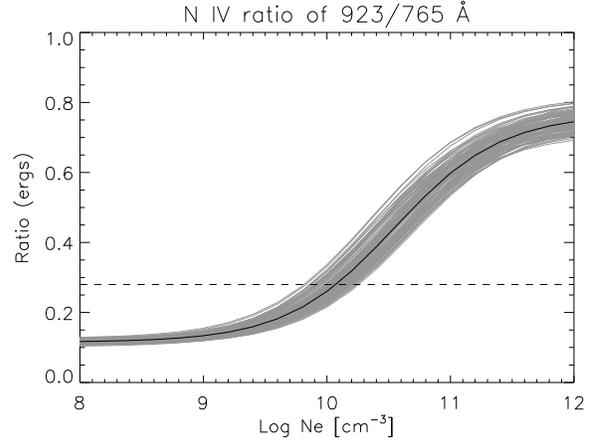, width=8.5cm,angle=0 }}
\caption{Main density diagnostic for \ion{N}{iv} at log $T$[K]=5. 
The dashed line indicates the quiet Sun observed value reported by Dufton et al (1979). }
\label{fig:ne}
\end{figure}

As reviewed in \cite{delzanna_mason:2018}, lines of Be-like ions 
have been extensively used in astrophysics to measure electron 
densities and temperatures. 
It is therefore useful to assess the impact of the uncertainties
in the atomic rates on the main ratios.

Similarly to the \ion{Fe}{xiii} study \citep{yu_etal:2018},
we have taken a `Monte Carlo' approach, i.e. 
we have calculated the level populations and line emissivities 
100  times by randomly varying each A-value and collisional rate
within some bounds, using the BSR values as a reference. 
To define the bounds for each rate, we have 
compared the three calculated values and taken the maximum 
relative deviation from the BSR values. We have limited the 
variation to a minimum of 2\% (as some variations are smaller than this) 
and a maximum of 80\%, neglecting the few order-of magnitude variations which,
as we have shown, have little effect on the line intensities.
Unlike \citep{yu_etal:2018}, where a normal distribution was adopted 
(with standard deviation equal to the bound), we have 
adopted a strict random distribution within the bounds.

\begin{figure}
\centerline{\epsfig{file=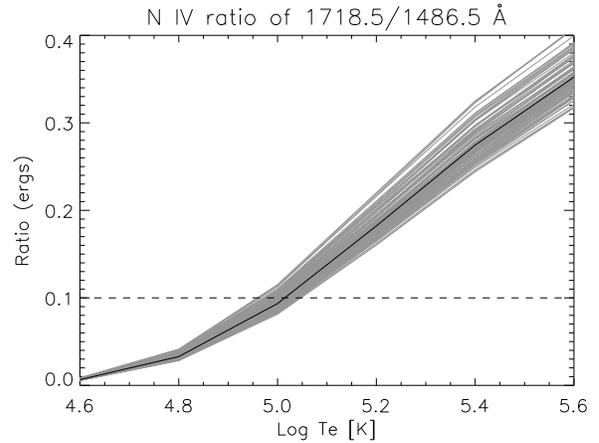, width=8.5cm,angle=0 }}
\caption{Main electron temperature diagnostic for \ion{N}{iv}, calculated
at   an electron density of  10$^{10}$ cm$^{-3}$.
The dashed line indicates the quiet Sun observed value 
reported by  Dufton et al (1979).
}
\label{fig:te}
\end{figure}

The main ratio to measure electron  densities is that 
of the multiplet of transitions from the 2p$^2$ $^3$P 
to the 2s2p $^3$P to the resonance line, the 
 2s$^2$ $^1$S$_{0}$--2s 2p $^1$P$_{1}$ at 765~\AA.
The multiplet of lines falls around 923~\AA\ and is blended with 
other transitions. 
 \cite{dufton_etal:1979} reported a deblended ratio of 
0.28 for the quiet Sun, and found an electron density of 1.5 $\times$ 10$^{10}$ cm$^{-3}$,
assuming a temperature of formation of log $T$[K]=5.1 and using the  
atomic data available at the time. 
Fig.~\ref{fig:ne} shows the theoretical ratio obtained from the BSR data
(black line) and the 99 random realisations (grey lines) for  
log $T$[K]=5.
The dashed line indicates the quiet Sun observed value 
reported by  \cite{dufton_etal:1979}. 
We can see that we obtain a similar value,
 about 1.2 $\times$ 10$^{10}$ cm$^{-3}$,  with an uncertainty of about 0.2 in dex.

The main ratio to measure electron temperatures 
is the  2s 2p $^1$P$_{1}$--2p$^2$ $^1$D$_{2}$ (1718.5~\AA)
vs. the 2s$^2$ $^1$S$_{0}$--2s 2p $^3$P$_{1}$ (1486.5~\AA).
Figure~\ref{fig:te} shows the the theoretical ratio obtained from the BSR data
(black line) and the 99 random realisations (grey lines), calculated for 
an electron density of $\times$ 10$^{10}$ cm$^{-3}$. 
The dashed line indicates the quiet Sun observed value 
reported by  \cite{dufton_etal:1979}. We can see that we obtain a 
temperature of log $T$[K]=5, the same value estimated by \cite{dufton_etal:1979}.
The uncertainty is about   0.05 in dex.

Clearly, a proper evaluation of the uncertainties should  also include 
the uncertainties in the observed values. Also, it should 
include a model of how the densities and temperatures
might vary along the line of sight, 
as the emissivities of the lines we have considered are dependent on 
both the electron densities and temperatures, to some degree. 
Such model would  depend critically on 
the source region observed.  As the  observations of the two ratios reported by
 \cite{dufton_etal:1979}
 were not simultaneous and were obtained in different 
conditions by different instruments, it is not possible to further 
explore this aspect in this example.

\section{Conclusions}

We have considered three independent calculations of the effective collision strengths
for \ion{N}{iv} 
which, for many transitions, show large order-of-magnitude discrepancies for many 
weak transitions and/or those involving high-lying levels. 

At first glance, such differences are of concern for  astrophysical 
applications, although 
we should point out that thay are inherent in $R$-matrix close-coupling methods based-on
truncated CI and/or CC expansions. 
We view such  discrepancies as an excellent way to provide a  
measure on the uncertainty in calculating rates for weak transitions and
to high-lying levels. 

Despite the differences, we have shown that in the case of 
\ion{N}{iv} excellent agreement (to within a relative 20\%) is found  
among the line intensities obtained from the three 
independent calculations considered here, for  the 
spectroscopically and astrophysically important emission lines.
Agreement in the line intensities obtained with the ICFT and the DARC effective
collision strengths  is even better, to within a relative 10\%.

The present  modelling  clearly shows that the few forbidden transitions where the 
 BSR, DARC and ICFT collision strengths  differ significantly are not really 
relevant for any astrophysical application where densities are below 10$^{15}$ cm$^{-3}$.
If low densities such in astrophysical nebulae are considered,  the discrepancies are even smaller,
because the level populations are driven solely by the excitation rates from the ground state,
as this is the only populated level (cf. Fig.~1).

We expect similar results for the other ions along the  Be-like sequence. 
The metastable levels become populated at increasingly higher densities 
along the sequence (with increasing atomic number). Therefore, we would expect 
for the higher Z elements better agreement in the line intensities calculated 
from the different codes for say solar densities (10$^{8}$--10$^{12}$ cm$^{-3}$).

We have used the differences obtained by the three sets of calculations
as a measure of the uncertainty in each of the radiative and 
collisional rates. With simple Monte Carlo simulations, we have shown how 
such uncertainties affect the main diagnostic applications 
for this ion, to measure electron densities and temperatures. 
We suggest that such an approach should be adopted when estimating uncertainties
on the theoretical ratios.

\section*{Acknowledgments}
GDZ acknowledges support from STFC (UK) through the 
University of Cambridge DAMTP astrophysics grant.
NRB acknowledges support from STFC (UK) through the
University of Strathclyde UK APAP Network grant ST/R000743/1.
We thank J. Mao for useful comments on the manuscript.

\bibliographystyle{mn2e}

\bibliography{ref.bib}

\end{document}